\documentclass[showpacs,preprintnumbers,prl,twocolumn]{revtex4}%
\usepackage{amssymb}
\usepackage{amsfonts}
\usepackage{amsmath}
\usepackage{graphicx}
\usepackage{dcolumn}
\usepackage{bm}%
\providecommand{\U}[1]{\protect\rule{.1in}{.1in}}
\begin{document}
\title{Noise Squeezing in a Nanomechanical Duffing Resonator}
\author{Ronen Almog}
\author{Stav Zaitsev}
\author{Oleg Shtempluck }
\author{Eyal Buks}
\affiliation{Department of Electrical Engineering, Technion, Haifa 32000 Israel}

\begin{abstract}
We study mechanical amplification and noise squeezing in a nonlinear
nanomechanical resonator driven by an intense pump near its dynamical
bifurcation point, namely, the onset of Duffing bistability. Phase sensitive
amplification is achieved by a homodyne detection scheme, where the
displacement detector's output, which has correlated spectrum around the pump
frequency, is down converted by mixing with a local oscillator operating at
the pump frequency with an adjustable phase. The down converted signal at the
mixer's output could be either amplified or deamplified, yielding noise
squeezing, depending on the local oscillator phase.

\end{abstract}
\pacs{87.80.Mj 05.45.-a}
\maketitle

Micro/Nanoelectromechanical resonators allow a variety of applications such as
sensing, switching, and filtering \cite{Roukes2000,Craighead2000,Cleland}.
Improvement of force detection sensitivity is highly desirable for enhancing
the performance of nanomechanical detectors. A possible technique to improve
signal to noise ratio in such devices is to implement an on-chip mechanical
amplification. Mechanical amplification and thermomechanical noise squeezing
in microresonators have been achieved before using parametric amplification
\cite{Rugar,Carr}. In the present work, we demonstrate a novel scheme for
mechanical amplification based on a bifurcating dynamical system, exploiting
its high sensitivity to fluctuations near its bifurcation point
\cite{Wiesenfeld,DykmanSNSP,byeb,krommer,Nori,Chan}. This amplification scheme
has been used lately for quantum measurement of superconducting qubits
\cite{Siddiqi}. In our case, we use the onset of bistability in a
nanomechanical Duffing resonator as the bifurcation point. In a Duffing
resonator, above some critical driving amplitude, the response becomes a
multi-valued function of frequency in some finite frequency range, and the
system becomes bistable with jump points in the frequency response
\cite{Nayfebook,Landau}. We employ this mechanism for the first time in
nanomechanical resonators to demonstrate experimentally, high signal gain,
phase sensitive amplification and noise squeezing. The system under study,
coined as NanoMechanical Bifurcation Amplifier (NMBA), consists of a nonlinear
doubly clamped nanomechanical PdAu beam, excited capacitively by an adjacent
gate electrode. An intense pump signal drives the resonator near the onset of
bistability, enabling amplification of a small signal in a narrow bandwidth.
In a previous work we have demonstrated high intermodulation gain occuring in
the same region \cite{Almog}.

The experimental setup is shown in Fig. 1. The resonator is excited by two
sources (pump and small test signal or noise) and its vibrations are detected
optically using a knife-edge technique \cite{KnifeEdge}. The device is located
at the focal point of a lensed fiber which is used to focus laser light at the
beam and to collect the reflected light back to the fiber and to a
photodetector (PD). The PD signal is amplified, mixed with a local oscillator
(LO), low pass filtered and measured by a spectrum analyzer. The resonator
length is $l=100%
\operatorname{\mu m}%
$, width $w=600%
\operatorname{nm}%
$, and thickness $t=250%
\operatorname{nm}%
$. The gap separating the doubly clamped beam and the stationary side
electrode is $d=4%
\operatorname{\mu m}%
$ wide. The device is fabricated using bulk nano-machining process together
with electron beam lithography \cite{ebmlr}.%
\begin{figure}
[ptb]
\begin{center}
\includegraphics[
height=2.4215in,
width=3.3313in
]%
{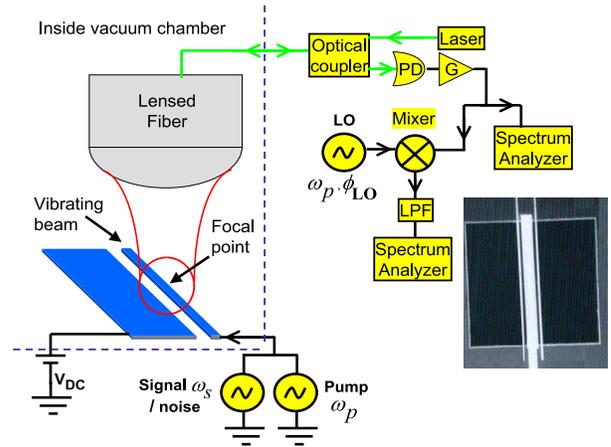}%
\caption{(Color online) The experimental setup. The device consisting of a
suspended doubly clamped nanomechanical resonator. The resonator is excited by
two phase locked sources (one source is used as a pump and the other one as a
small test signal or as a noise source). The resonator's vibrations are
detected optically. The inset shows an electron micrograph of the device.}%
\end{center}
\end{figure}

The nonlinear dynamics of the fundamental mode of a doubly clamped beam driven
by an external force per unit mass $F(t)$ can be described by a Duffing
oscillator equation for a single degree of freedom $x$
\begin{equation}
\ddot{x}+2\mu(1+\beta x^{2})\dot{x}+\omega_{0}^{2}(1+\kappa x^{2})x=F(t),
\label{eom}%
\end{equation}
where $\mu$ and $\beta$ are the linear and nonlinear damping constants
respectively, $\omega_{0}/2\pi\simeq500%
\operatorname{kHz}%
$ is the resonance frequency of the fundamental mode, and $\kappa$ is the
cubic nonlinear constant. Generally, for resonators driven using a bias
voltage applied to a side electrode, Eq. (\ref{eom}) should contain additional
parametric terms \cite{Rugar,Blencowe}. In our case however, the prefactors of
these parametric terms are at least one order smaller below threshold and thus negligible.

The relative importance of nonlinear damping can be characterized by the
dimensionless parameter $p=2\sqrt{3}\mu\beta/\kappa\omega_{0}$ \cite{Stav}.
The relatively low value of $p\simeq0.05,$ obtained from the measured values
of $\omega_{0}$, $\mu$ and the frequency of the bifurcation point \cite{Stav},
indicates that the effect of nonlinear damping in the present case is
relatively weak. To investigate nonlinear amplification of a small test
signal, the resonator is driven by an applied force $F(t)=f_{p}\cos(\omega
_{p}t)+f_{s}\cos(\omega_{s}t+\varphi),$ composed of an intense \textit{pump}
with frequency $\omega_{p}=\omega_{0}+\sigma,$ amplitude $f_{p},$ and a small
force (called \textit{signal}) with frequency $\omega_{s}=\omega_{p}+\delta,$
relative phase $\varphi$, and amplitude $f_{s}$, where $f_{s}\ll f_{p}$ and
$\sigma,\delta\ll\omega_{0}.$ This is achieved by applying a voltage of the
form $V=V_{dc}+v_{p}\cos(\omega_{p}t)+v_{s}\cos(\omega_{s}t+\varphi)$ where
$V_{dc}$ is a dc bias (employed for tuning the resonance frequency) and
$v_{s}\ll v_{p}\ll V_{dc}$. The resonator's displacement has spectral
components at $\omega_{p},$ $\omega_{s},$ and at the intermodulations
$\omega_{p}\pm k\delta$ where $k$ is an integer. The one at frequency
$\omega_{i}=\omega_{p}-\delta$ is called the \textit{idler} component.

Strong correlation between the signal and the idler, occurring near the edge
of the bistability region, could be exploited for phase sensitive
amplification and noise squeezing \cite{byeb,Yurke2}. This is achieved by a
homodyne detection scheme, where the displacement detector's output is down
converted by mixing with a LO operating at frequency $\omega_{p}$ with an
adjustable phase $\phi_{\mathrm{LO}}$ and phase locked to the pump. The
mixer's output (IF port) has a spectral component at frequency $\delta,$ which
is proportional to the phasor sum of the signal and the idler, yielding phase
sensitive amplification, controlled by $\phi_{\mathrm{LO}}$. In the notation
of Ref. \cite{Almog}, the displacement $x(t)$ is given by
\begin{equation}
x(t)=\frac{1}{2}A(t)e^{i\omega_{p}t}+c.c.\ ,
\end{equation}
where $A(t)=a_{p}+a_{s}e^{i\delta t}+a_{i}e^{-i\delta t}$ is a slowly varying
function (relative to the time scale $1/\omega_{p})$ and the complex numbers
$a_{p,}$ $a_{s}$ and $a_{i}$ are the pump, signal and idler spectral
components of $A(t)$ respectively. Suppose that the LO voltage is given by
$V^{\mathrm{LO}}(t)=V_{0}^{\mathrm{LO}}\cos(\omega_{p}t+\phi_{\mathrm{LO}})$
and the mixer's output is given by $V_{\mathrm{MO}}=Mx(t)V^{\mathrm{LO}}(t)$
where $M$ is a constant term depending on the optical detector's sensitivity,
amplification and the mixing factor. After passing through a low pass filter
(LPF), the output signal is%
\[
\frac{1}{4}MV_{0}^{\mathrm{LO}}[A(t)e^{-i\phi_{\mathrm{LO}}}+c.c.]\ .
\]

The measured quantity is the amplitude $R(\delta)$ of the spectral component
of the output signal at frequency $\delta$. $R(\delta)$ depends on the LO
phase $\phi_{LO}$ and is given by%
\begin{equation}
R(\delta)=\frac{1}{2}MV_{0}^{\mathrm{LO}}\left\vert a_{s}e^{-i\phi
_{\mathrm{LO}}}+a_{i}^{\ast}e^{i\phi_{\mathrm{LO}}}\right\vert \ .
\end{equation}
As $\phi_{\mathrm{LO}}$ is varied, the term $\left\vert a_{s}e^{-i\phi
_{\mathrm{LO}}}+a_{i}^{\ast}e^{i\phi_{\mathrm{LO}}}\right\vert $ oscillates
between the minimum value $\left\vert \left\vert a_{s}\right\vert -\left\vert
a_{i}\right\vert \right\vert $ and the maximum one $\left\vert a_{s}%
\right\vert +\left\vert a_{i}\right\vert $. When $\delta\rightarrow0,$ and
$a_{p}$ is tuned to the bifurcation point, we have $\left\vert a_{s}%
\right\vert =\left\vert a_{i}\right\vert \simeq f_{s}/2\omega_{0}\delta$
\cite{Almog}, hence $R(\delta)_{\max}=MV_{0}^{LO}f_{s}/2\omega_{0}\delta$ and
$R(\delta)_{\min}/R(\delta)_{\max}\rightarrow0.$ The factor $\ \Delta\equiv
R(\delta)_{\max}-R(\delta)_{\min}$\ characterizes the phase dependence of the
amplification. An example of a measurement of $R(\delta)$ vs. $\phi_{LO}$ is
shown in Fig. 3(a).

To study the response to injected noise, the resonator is excited by a fixed
pump near the bifurcation point, together with white noise. In this case Eq.
(\ref{eom}) is a Langevin equation with $F(t)=f_{p}\cos(\omega_{p}t)+F_{n}(t)
$ where $F_{\mathrm{n}}(t)$ is a white noise having a vanishing mean
$\left\langle F_{\mathrm{n}}(t)\right\rangle =0$, and spectral density
$S_{F_{n}}=4\omega_{0}k_{B}T_{\mathrm{eq}}/mQ$ \cite{Cleland}. Here
$T_{\mathrm{eq}}$ is the equivalent temperature of the applied voltage noise,
$m$ is the effective mass of the fundamental mode. In this case, the
displacement spectral density measured at the mixer's output will consist of
two contributions, namely, the pump response ($\delta$-function peaked at
$\delta=0$), and a continuous part $S_{x}(\delta)$ due to noise. In the limit
$\delta\rightarrow0$ the spectral density $S_{x}$, which was calculated in
Ref. \cite{BYmassDetection}, is given by%

\begin{equation}
S_{x}=\frac{1+2\zeta\cos(\phi_{\mathrm{LO}}-\phi_{0})+\zeta^{2}}{(1-\zeta
^{2})^{2}}S_{x0}\ , \label{NoiseR}%
\end{equation}
where
\[
S_{x0}=S_{F_{n}}/\left\{  4\omega_{0}^{2}\left[  \mu^{2}+\left(  \omega
_{p}-\omega_{0}-\frac{3}{2}\omega_{0}\kappa\left\vert a_{p}\right\vert
^{2}\right)  ^{2}\right]  \right\}
\]
and $\phi_{0}$ and $\zeta$ are real parameters. While $\zeta$ vanishes in the
linear region, its largest value $\zeta=1$ is obtained along the edge of the
bistability region. Eq. (\ref{NoiseR}) implies that when $\delta\rightarrow0$,
the output noise will oscillate between a maximum value, corresponding to the
amplified quadrature, and a minimum one, corresponding to the deamplified (or
squeezed) quadrature, as $\phi_{\mathrm{LO}}$ is varied \cite{BYmassDetection}%
\[
\left[  S_{x}\right]  _{\max}=\frac{S_{x0}}{(1-\zeta)^{2}};\left[
S_{x}\right]  _{\min}=\frac{S_{x0}}{(1+\zeta)^{2}}\ .
\]
Thus, the largest amplification obtained by this model diverges at the
bifurcation point, whereas noise squeezing is limited to a factor of 4.%

\begin{figure}
[ptb]
\begin{center}
\includegraphics[
trim=0.000000in 0.000000in -0.069611in -0.377198in,
height=2.5045in,
width=3.3321in
]%
{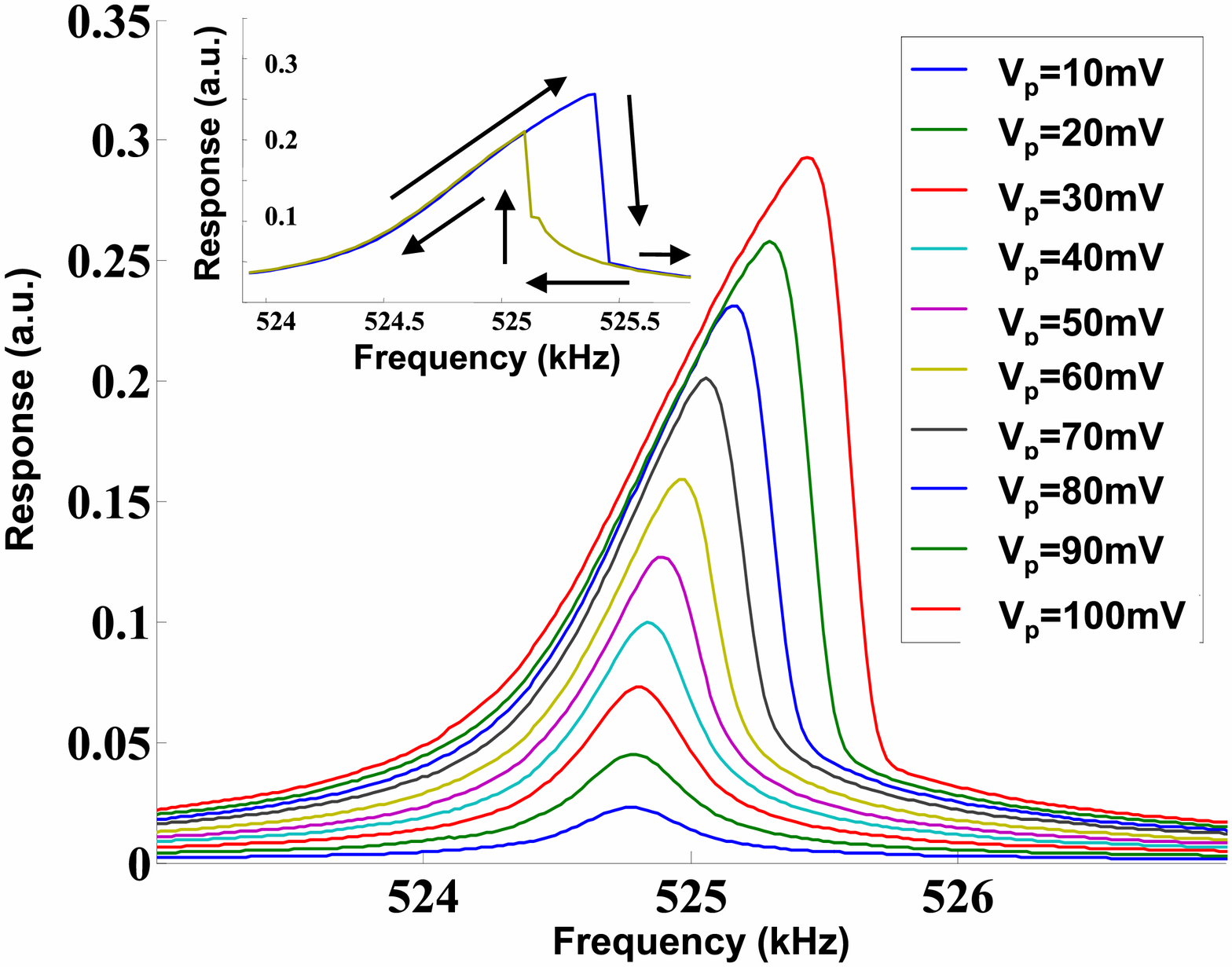}%
\caption{(Color online) Typical frequency response curves for various
excitation voltages $Vp$. The inset shows hysteresis response for
$Vp=90\operatorname{mV}.$}%
\end{center}
\end{figure}

We now turn to describe the experimental steps. As a first step we find the
onset of bistability and characterize the bistability region. This is achieved
by sweeping the pump frequency upward and back downward for different constant
excitation amplitudes, without additional small signal or noise. Typical
response curves are shown in Fig. 2. The bistability region and the
bifurcation point \textrm{B}$_{\mathrm{p}}$ (marked with a circle) are shown
in Fig. 3(b). In the next step, we characterize small signal amplification by
exciting the resonator with pump and small test signal where $v_{p}/v_{s}=25$.
Measurements of $\Delta$ vs. frequency are shown in Fig. 3(c) for four pump
amplitudes (related to lines (1)-(4) in Fig. 3(b)). The response of the
frequency upward (downward) sweep is depicted with black (green) line. For
$v_{p}=50%
\operatorname{mV}%
$ (Fig. 3(c)-1) the frequency sweep is contained within the monostable region
and consequently the value of $\Delta$ is relatively small. For $v_{p}=70%
\operatorname{mV}%
$ (Fig. 3(c)-2), $v_{p}=90%
\operatorname{mV}%
$ (Fig. 3(c)-3), and $v_{p}=110%
\operatorname{mV}%
$ (Fig 3(c)-4), on the other hand, the frequency sweeps cross the bistability
region and two peaks are seen for $\Delta$, corresponding to the upward and
downward frequency sweeps. These peaks originate from the high signal
amplification in the jump points of the pump response. Note that in this case
the width of the hysteresis loop (which is the distance between the peaks) is
smaller relative to the case when the pump is the only excitation.%

\begin{figure}
[h]
\begin{center}
\includegraphics[
height=5.7009in,
width=3.4938in
]%
{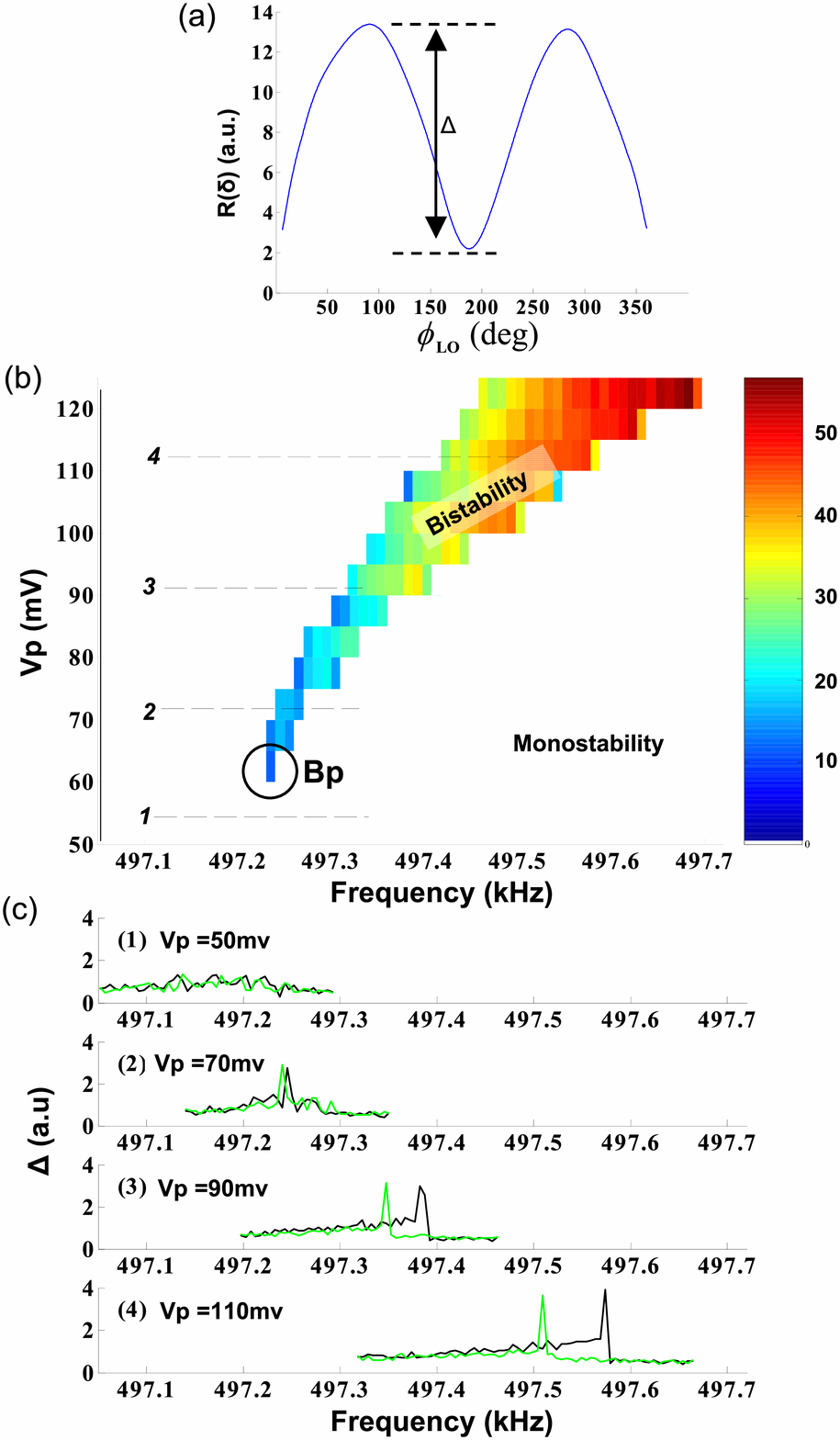}%
\caption{(Color online) (a) Measured $R(\delta)$ vs. LO phase $\phi
_{\mathrm{LO}}$. (b) Measurement of the bistability (hysteresis) region. The
bifurcation point \textrm{B}$_{\mathrm{p}}$ is marked with a circle. (c) The
parameter $\Delta$ vs. frequency for four different $Vp$ values (related to
lines (1)-(4) in Fig. 3(b)).}%
\end{center}
\end{figure}

We now turn to investigate the resonator response to pump and noise. First,
the bifurcation point (\textrm{B}$_{\mathrm{p}}$) is located. A frequency
response of the beam, excited by the pump (without noise) in the vicinity of
\textrm{B}$_{\mathrm{p}}$ is shown in Fig. 4(a). In the next step, the pump
frequency is fixed to the bifurcation point and we add white noise to the
excitation having spectral density $S_{\text{Vnoise}}^{1/2}=$1$%
\operatorname{mV}%
$/$\sqrt{%
\operatorname{Hz}%
}$.

The measured spectrum taken around the pump frequency (see Fig. 4(b))
demonstrates strong amplification occurring in this region, a manifestation of
the noise rise phenomenon \cite{Bryant}. There is a good agreement between the
theoretical fit ($\delta^{-1}$ dependence) \cite{byeb} to the experimental
data for $\delta>50%
\operatorname{Hz}%
$. For smaller values of $\delta$ the model breaks down due to high order terms.%

\begin{figure}
[ptb]
\begin{center}
\includegraphics[
trim=0.000000in 0.000000in 0.000000in 0.412203in,
height=4.3474in,
width=3.3096in
]%
{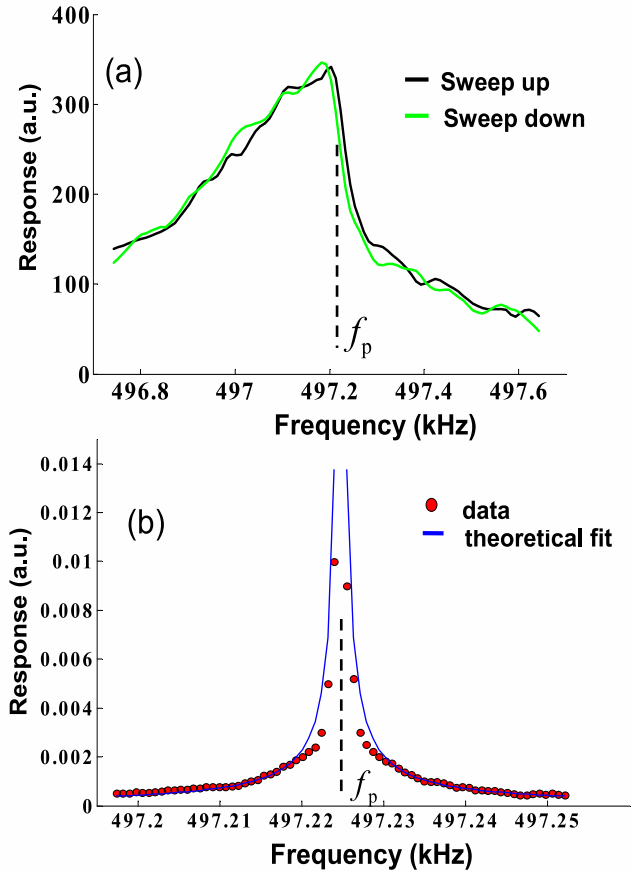}%
\caption{(Color online) (a) Pump response near \textrm{B}$_{\mathrm{p}}$.
Upward and downward sweeps are seen in black and green respectively. (b)
Averaged spectrum response for pump and noise excitation. The input noise
spectral density is 1$\operatorname{mV}$/$\sqrt{\operatorname{Hz}}$. Circles
indicate the experimental data, whereas a theoretical fit is seen as a blue
line.}%
\end{center}
\end{figure}

Noise squeezing is demonstrated in Fig. 5 where $\delta=10%
\operatorname{Hz}%
,$ $S_{Vnoise}^{1/2}=$1$%
\operatorname{mV}%
$/$\sqrt{%
\operatorname{Hz}%
}$ and $S_{x}^{1/2}$ is plotted vs. the LO phase $\phi_{\mathrm{LO}}$. Here
the sweep time is $6%
\operatorname{s}%
$ and the resolution bandwidth is $2%
\operatorname{Hz}%
.$ The blue line demonstrates the case where the pump is in the vicinity of
the bifurcation point, whereas the green line demonstrates the case where the
pump is far from the bifurcation point. The noise amplitude amplification is
about 6. The deamplified (squeezed) quadrature is below the measurement noise
floor, hence it can't be measured.%

\begin{figure}
[ptb]
\begin{center}
\includegraphics[
trim=0.000000in 0.000000in 0.268265in 0.000000in,
height=2.5036in,
width=3.371in
]%
{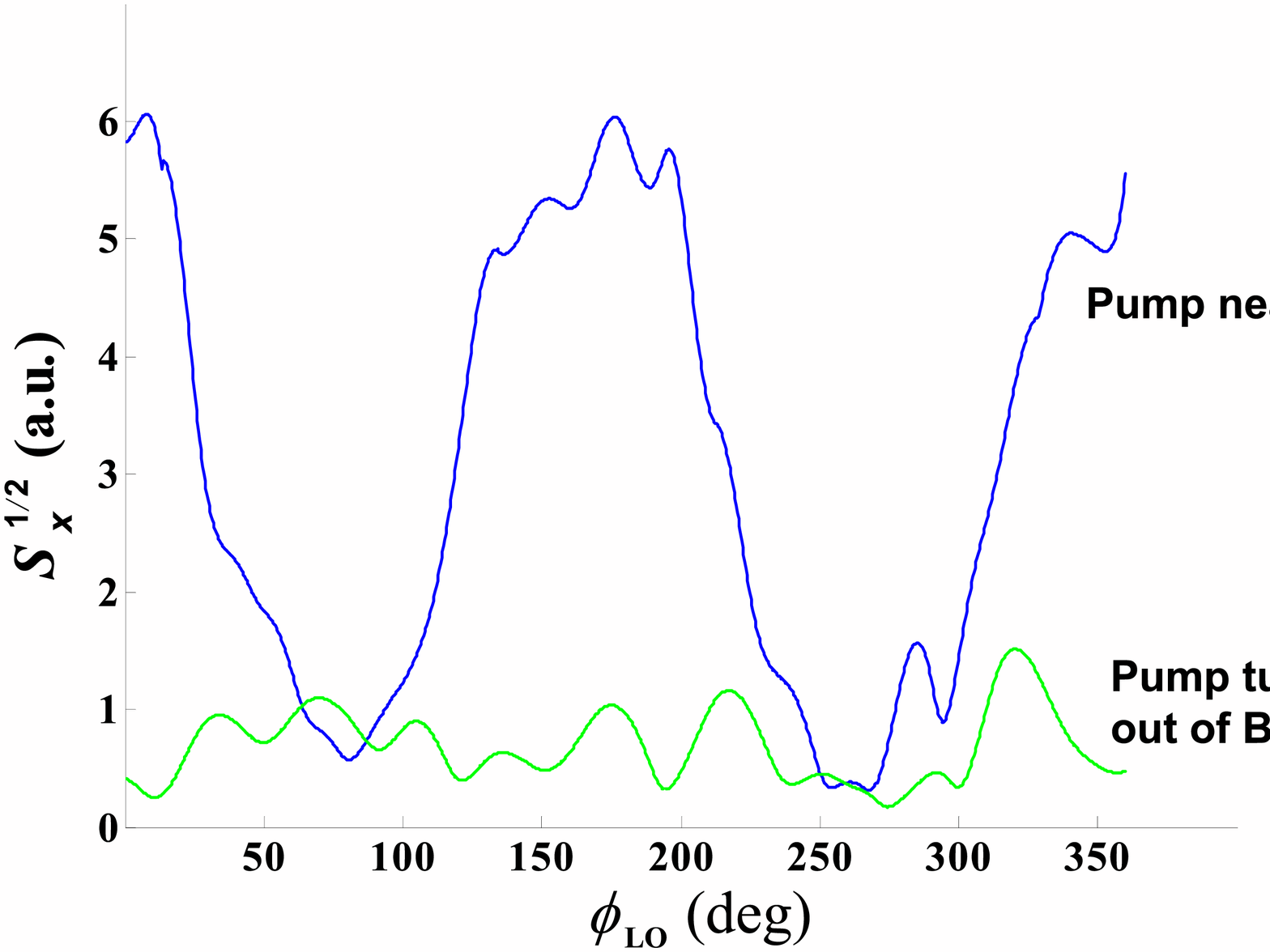}%
\caption{(Color online) Noise squeezing. The spectral component $S_{x}^{1/2}$
vs. $\phi_{\mathrm{LO}}$ for $\delta=10\operatorname{Hz}$. The resonator is
excited by pump and noise. Blue line - pump near Bp, green line - pump tuned
out of Bp (200Hz higher).}%
\end{center}
\end{figure}

To summarize, the nonlinear region of nanomechanical resonators could be
exploited for both signal amplification and noise reduction which could be
useful for detection of weak forces. A possible application for our noise
squeezing scheme is sensitive mass detection \cite{BYmassDetection}, which can
be achieved by operating close to the bifurcation point and adjusting
$\phi_{\mathrm{LO}}$ to maximize the mass sensitivity.

This work is supported by the Israeli ministry of science, Intel Corp.,
Israel-US binational science foundation, and by Henry Gutwirth foundation. The
authors wish to thank Bernard Yurke for helpful discussions.

\end{document}